\documentclass[lettersize,journal]{IEEEtran}
\usepackage{amsmath,amsfonts}
\usepackage{algorithmic}
\usepackage{algorithm}
\usepackage{array}
\usepackage[caption=false,font=normalsize,labelfont=sf,textfont=sf]{subfig}
\usepackage{textcomp}
\usepackage{stfloats}
\usepackage{url}
\usepackage{verbatim}
\usepackage{graphicx}
\usepackage{cite}
\usepackage{booktabs}
\usepackage{multirow}
\usepackage{color}
\usepackage{xcolor}
\definecolor{lightred}{rgb}{1.0, 0.5, 0.5} 
\usepackage{hyperref}       

\begin{document}

\title{PSCodec: A Series of High-Fidelity Low-bitrate Neural Speech Codecs Leveraging Prompt Encoders}

\author{Yu Pan, Xiang Zhang, Yuguang Yang, Jixun Yao, Yanni Hu, Jianhao Ye, \\ Hongbin Zhou,  Lei Ma, Jianjun Zhao
\thanks{Yu Pan and Jiajun Zhao are with the School of Information Science and Electrical Engineering, Kyushu University, Fukuoka 8190385, Japan (e-mail: panyu.ztj@gmail.com; zhao@ait.kyushu-u.ac.jp.)}
\thanks{Xiang Zhang, Yuguang Yang, Jixun Yao, Yanni Hu, Jianhao Ye, Hongbin Zhou are with the Shanghai Ximalaya Technology Co Ltd, Shanghai 201203, China (e-mail: xiang2.zhang@ximalaya.com; yuguang.yang@ximalaya.com; yaoxunji@gmail.com; yanni.hu@ximalaya.com; jianhao.ye@ximalaya.com; hongbin.zhou@ximalaya.com.)}
\thanks{Lei Ma is with the Department of Computer Science, The University of Tokyo, Tokyo 1138656, Japan, and the Department of Electrical and Computer Engineering, University of Alberta, Edmonton T2G 1S6, Canada (e-mail: ma.lei@acm.org.)}
\thanks{Yu Pan and Xiang Zhang contributed equally to this work.}
}


\markboth{Journal of \LaTeX\ Class Files,~Vol.~14, No.~8, August~2021}%
{Shell \MakeLowercase{\textit{et al.}}: A Sample Article Using IEEEtran.cls for IEEE Journals}

\maketitle

\begin{abstract}
Neural speech codecs have recently emerged as a focal point in the fields of speech compression and generation.
Despite this progress, achieving high-quality speech reconstruction under low-bitrate scenarios remains a significant challenge. 
In this paper, we propose \emph{PSCodec}, a series of neural speech codecs integrated effective prompt encoders, including \emph{PSCodec-Base}, \emph{PSCodec-DRL-ICT}, and \emph{PSCodec-CasAN}, which excel in delivering high-performance speech reconstruction with low bandwidths.
Specifically, we first introduce \emph{PSCodec-Base}, which leverages a pretrained speaker verification model-based prompt encoder (\emph{VPP-Enc}) and a learnable Mel-spectrogram-based prompt encoder (\emph{MelP-Enc}) to effectively disentangle and integrate voiceprint and Mel-related features in utterances. 
To further enhance feature utilization efficiency, we propose \emph{PSCodec-DRL-ICT}, incorporating a structural similarity (SSIM) based disentangled representation loss (DRL) and an incremental continuous training (ICT) strategy. 
While \emph{PSCodec-DRL-ICT} achieves impressive performance, its reliance on extensive hyperparameter tuning and multi-stage training introduces complexity. 
To circumvent these limitations, we propose \emph{PSCodec-CasAN}, utilizing an advanced cascaded attention network (CasAN) to enhance representational capacity of the entire system.
Extensive experiments indicate that our proposed \emph{PSCodec-Base}, \emph{PSCodec-DRL-ICT}, and \emph{PSCodec-CasAN} all significantly outperform several state-of-the-art neural codecs, offering substantial improvements in speech reconstruction quality and speaker similarity under low-bitrate conditions.
\end{abstract}


\section{Introduction}
\IEEEPARstart{N}{eural} speech codecs, 
which compress utterances into low-dimensional discrete tokens and subsequently decode them for reconstruction, have garnered considerable attention in the domains of speech enhancement \cite{xue2024low,chiang2024restorative,buthe2024nolace} and speech generation \cite{wang2023neural,betker2023better,borsos2023audiolm,zhang2023speak,wang2023lm,martin2024enhancing,yang2024takin,chen2024takin}.

\begin{figure}
\centering
	\includegraphics[height=4cm,width=!]{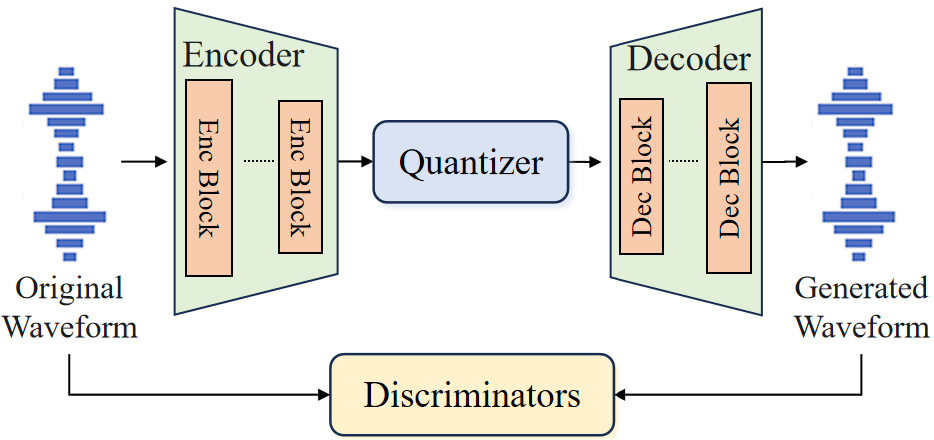}
	\caption{The schematic of conventional neural speech codecs.}
        \label{fig:test}
\end{figure}

In essence, the neural speech codec can be conceptualized as a technique for lossy compression, aiming to minimize the bitrate while ensuring that the quality of the reconstructed speech remains largely preserved, thereby facilitating the efficient storage, transmission, and utilization of utterances.
In recent times, amidst the swift progress of deep learning, impressive advancements have been made in the field of neural speech codecs.
Generally speaking, current state-of-the-art (SOTA) neural speech codecs \cite{garbacea2019low,zeghidour2021soundstream,defossez2022high,yang2023hifi,jenrungrot2023lmcodec,du2024funcodec} typically employ a straightforward Encoder-Quantizer-Decoder workflow to achieve speech compression and reconstruction in an end-to-end (E2E) manner, as shown in Fig. \ref{fig:test}. 
Under this paradigm, 1) the Encoder initially transforms the input speech from the time or frequency domain into compressed representational spaces via deep neural networks; 
2) the Quantizer approximates the compressed features by assigning them to the closest features within the learnable codebooks; 
3) quantized features are subsequently fed into the Decoder for upsampling, with the aim of reconstructing the original input waveform.
Although these approaches manifest improved performance over earlier techniques \cite{dietz2015overview,valin2016high,klejsa2019high}, they encounter a significant challenge due to the reliance on multiple codebooks or excessively long token sequences.

To this end, numerous studies have sought to enhance speech reconstruction at lower bitrates.
For instance, \cite{jenrungrot2023lmcodec} introduced a causal convolutional codec that encodes audio into a hierarchy of coarse-to-fine tokens using residual vector quantization (RVQ). 
\cite{kumar2024high} integrated improved vector quantization techniques with adversarial losses to sustain speech quality at low bandwidths. However, due to the limited representational capacity of the encoder, these methods continue to struggle in maintaining the quality of encoded features at higher compression rates.
Recent efforts by \cite{yang2024generative,ju2024naturalspeech,san2024discrete} have leveraged latent diffusion models \cite{ho2020denoising,rombach2022high} to achieve high-quality speech reconstruction at low bitrates. Nevertheless, these approaches are constrained by the inherent computational overhead of diffusion models, leading to suboptimal real-time performance.
Conversely, \cite{ren2023fewer} proposed TiCodec, which encodes and quantizes time-invariant elements from speech into a separate code. Inspired by this work, \cite{li2024singlecodec} introduced Single-Codec, which employs a disentangled VQ-VAE to separate speech into a phonetically rich discrete sequence and a time-invariant embedding. Despite this innovation, Single-Codec relies on Mel-Spectrogram (Mel) reconstruction using BigVGAN \cite{lee2023bigvgan}, which inevitably results in partial information loss and compromises its overall performance.
Hence, based on the aforementioned analysis, the pursuit of high-fidelity neural speech codecs under low bitrate scenarios remains an ongoing challenge.

To address these problems, this paper introduces \emph{PSCodec}, a series of scalable neural speech codecs leveraging advanced prompt encoders: \emph{PSCodec-Base}, \emph{PSCodec-DRL-ICT}, and \emph{PSCodec-CasAN}, to achieve high-quality speech reconstruction with low bandwidths.
\textbf{Essentially, these systems emphasize the efficient decoupling and exploitation of specific speech attribute features to enhance overall performance.}
To elaborate, building upon a HifiCodec-like \cite{yang2023hifi} architecture, we begin by proposing \emph{PSCodec-Base} that integrates two specialized prompt encoders: VPP-Enc and MelP-Enc.
The VPP-Enc is a pre-trained speaker verification (SV) model-based prompt encoder, while the MelP-Enc is a learnable Mel-based prompt encoder. 
Together, these encoders can effectively disentangle voice-print (VP) characteristics and Mel-correlated information—such as emotion, intonation, and accent from human speech, thereby improving both compression efficiency and reconstruction quality.
Nonetheless, relying exclusively on prompt encoders does not yield optimal results, as challenges persist in maximizing the efficiency of individual encoders and improving the overall representational capacity of the system. 
Accordingly, we introduce \emph{PSCodec-DRL-ICT}, a neural speech codec framework that leverages a structural similarity (SSIM) \cite{wang2004image} based disentangled representation loss (DRL) and an incremental continuous training (ICT) approach.
While \emph{PSCodec-DRL-ICT} excels in speech reconstruction by achieving stable training and minimizing redundancy across encoded features, its reliance on extensive hyperparameter tuning and multi-stage training introduces extra complexity and labor requirements.
To mitigate these constraints, we further propose \emph{PSCodec-CasAN}, which integrates an advanced cascaded attention network (CasAN)  to enable the adaptive augmentation of the system’s overall representational capacity, thus enhancing its final performance.

Experimental results indicate that our \emph{PSCodec-Base}, \emph{PSCodec-DRL-ICT} and \emph{PSCodec-CasAN} systems can delivery high-performance speech reconstruction at various low bitrates, outperforming several well-known SOTA neural speech codecs, e.g., Encodec \cite{defossez2022high}, AudioDec \cite{wu2023audiodec}, HifiCodec \cite{yang2023hifi}, TiCodec \cite{ren2023fewer},  APCodec \cite{ai2024apcodec}, and DAC \cite{kumar2024high}.
Particularly, at a bitrate of 675 bps with a single codebook, the proposed \emph{PSCodec-DRL-ICT} achieves 18.1\%, 3.1\%, 46.6\%, 8.3\%, 225.5\%, 13.9\%, and 2.3\% relative improvements over the second-best HifiCodec in terms of perceptual evaluation of speech quality (PESQ) \cite{rix2001perceptual}, short-time objective intelligibility (STOI) \cite{taal2010short}, virtual speech quality objective listener\footnote{\url{https://github.com/google/visqol}} (ViSQOL) \cite{chinen2020visqol}, UTMOS\footnote{\url{https://github.com/tarepan/SpeechMOS}} \cite{saeki2022utmos}, scale-invariant signal-to-noise ratio (SiSNR), mel-cepstrum distortion (MCD), and speaker embedding cosine similarity (SECS), on the LibriTTS test-clean set, highlighting the superiority of our proposed PSCodec-based framework.

In summary, this study presents the following contributions:
\begin{itemize}
\item We propose \emph{PSCodec}, a series of prompt encoders-based neural speech codecs that all achieve high-performance speech reconstruction with low bandwidths.
\item We introduce \emph{PSCodec-Base} that comprises two specialized prompt encoders to decouple and integrate VP and Mel-associated elements within utterances, thereby enhancing its compression rate and reconstruction performance.
\item Based on \emph{PSCodec-Base}, we propose \emph{PSCodec-DRL-ICT} that leverages a SSIM-based DRL alongside an ICT approach to improve its overall feature utilization efficiency.
\item Building upon \emph{PSCodec-Base}, we present \emph{PSCodec-CasAN} that incorporates an advanced CasAN to enhance the overall representational capacity of the entire system.
\item Experiments show that our proposed systems all surpass SOTA neural codecs across all metrics, both in-domain (ID) and out-of-domain (OOD), at various low bitrate scenarios, showcasing their effectiveness, robustness, and superiority.
\end{itemize}

\begin{figure*}[htbp]
\centering
	\includegraphics[height=6.4cm,width=!]{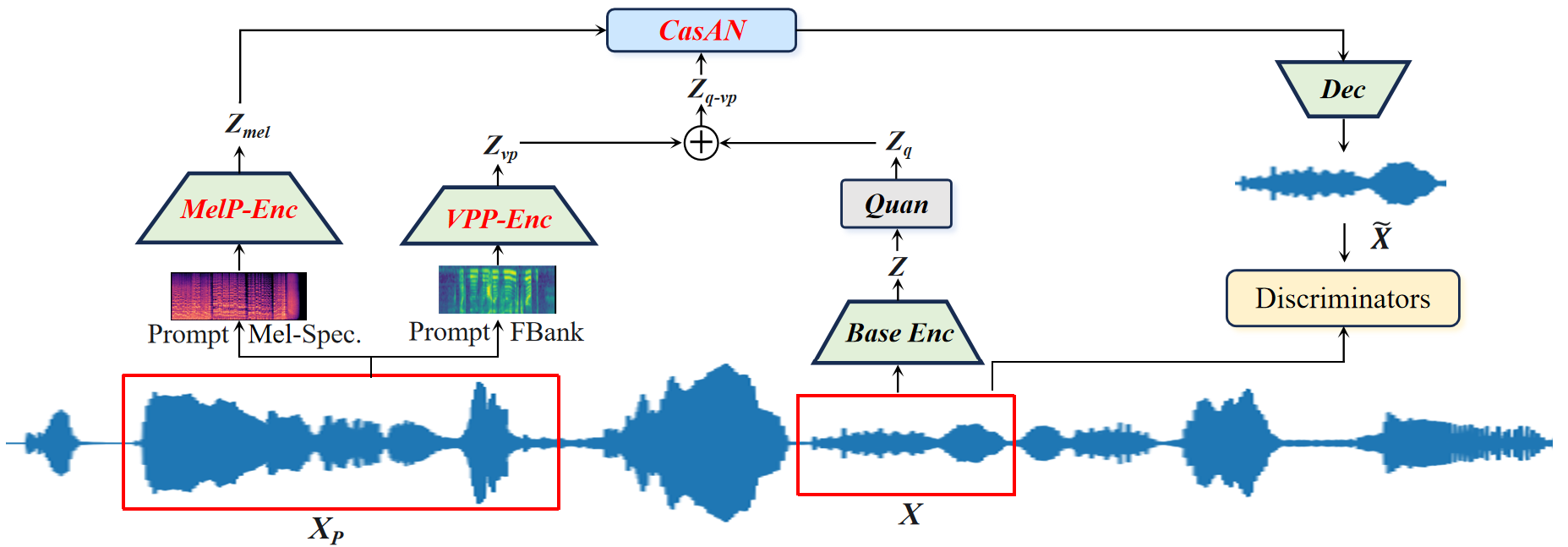}
	\caption{The schematic of the proposed PSCodec-CasAN training framework. The CasAN denotes our proposed cascaded attention network.}
        \label{fig:PSCodec-casan}
\end{figure*}

The remainder of this work is organized as follows. Section \ref{sec:Related Work} presents an overview of the related work. In Section \ref{sec:METHODOLOGY}, we detail the proposed \emph{PSCodec-Base}, \emph{PSCodec-DRL-ICT}, and \emph{PSCodec-CasAN} systems, with  their architectures shown in Fig. \ref{fig:PSCodec-base}, Fig. \ref{fig:PSCodec-DRL-ICT}, and Fig. \ref{fig:PSCodec-casan}, respectively. 
Section \ref{sec:EXPERIMENTS} outlines the experimental setup, results, and analysis. Finally, Section \ref{sec:CONCLUSIONS} concludes the paper.

\section{Related Work}
\label{sec:Related Work}

\subsection{Discrete Neural Speech Codec.}
Recently, E2E neural speech codecs have emerged as the leading solution for speech compression and reconstruction, garnering great attention in downstream fields such as speech generation and enhancement \cite{xue2024low,chiang2024restorative,buthe2024nolace,wang2023neural,betker2023better,borsos2023audiolm,zhang2023speak,wang2023lm,martin2024enhancing,yang2024takin,chen2024takin}. 
The pioneering work of \cite{garbacea2019low} introduced the first E2E neural codec model based on VQ-VAE \cite{van2017neural} and a WaveNet \cite{van2016wavenet} decoder, achieving superior performance at a 1.6 kbps bitrate. 
\cite{zeghidour2021soundstream} presented a novel framework composed of encoder, decoder, and residual vector quantizer (RVQ) termed SoundStream, achieving impressive performance over a wide range of bitrates.
Afterwards, \cite{defossez2022high} and \cite{yang2023hifi} advocated Encodec and HifiCodec, respectively, which have been extensively adopted. 
Nevertheless, these methods exhibit performance degradation at lower bitrates, posing challenges for high-fidelity speech reconstruction and downstream speech generation systems. 
To alleviate these issues, various approaches have been introduced, focusing on optimizing model architectures \cite{jenrungrot2023lmcodec,kumar2024high,san2024discrete}, decomposing speech representations \cite{ren2023fewer,jiang2023disentangled,li2024singlecodec}, and employing advanced diffusion models \cite{yang2024generative,ju2024naturalspeech,san2024discrete} to enhance reconstruction quality at low bitrates. 
Despite these notable advancements, there remains considerable potential for further improvement in terms of training efficiency and overall performance under low-bitrate scenarios.

\begin{figure*}[htbp]
\centering
	\includegraphics[height=6.6cm,width=!]{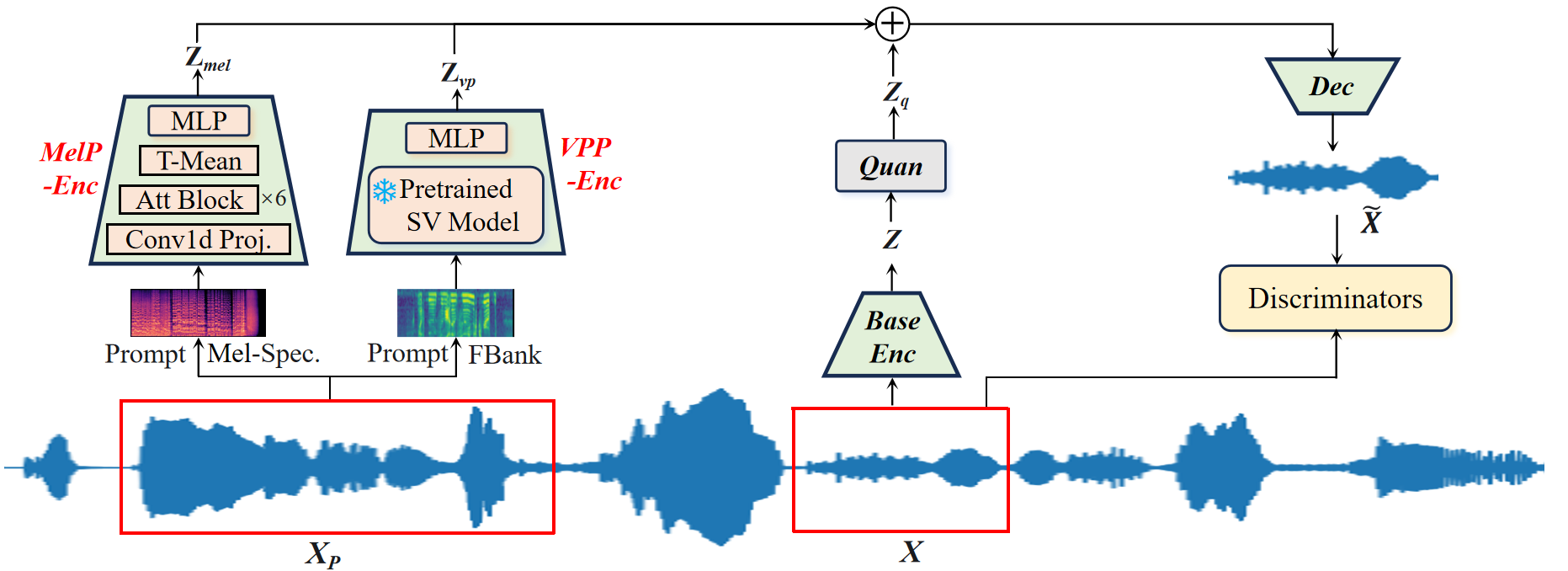}
        \caption{The architecture of our proposed PSCodec-Base training framework, comprising five fundamental components: the base encoder (Base Enc), quantizer (Quan), decoder (Dec), VPP-Enc, and MelP-Enc. Here, $X$, $X_P$, and $\tilde{X}$ represent the input speech, input prompt, and reconstructed waveform, respectively.}
        \label{fig:PSCodec-base}
\end{figure*}

\subsection{Disentangled Representation Learning.}
Disentangled representation learning, crucial in machine and deep learning, has been widely employed in the area of speech processing \cite{chen2021again,9747763,9746973,pan2023gemo, yao2023promptvc, pan2024gmp}. 
At its core, disentangled representation learning aims to decompose complex data into distinct latent spaces, with each space representing an independent factor or attribute. This structured decomposition not only enables a deeper understanding of the underlying data characteristics but also facilitates more effective manipulation and control of these attributes.
For example,
\cite{9746973} advocated an unsupervised loss function, extending MixIT with speech recognition embedding and disentanglement loss, to resolve domain mismatches and balance speech enhancement with ASR performance.
In the voice conversion task, several studies have endeavored to integrate diverse modules, including adaptive instance normalization \cite{chen2021again} and mutual information loss \cite{wang2021vqmivc}, aimed at disentangling linguistic content from speaker timbre features.
Despite significant progressions, the inherent complexity of speech signals—characterized by the coexistence of overlapping attributes such as phonetic, prosodic, emotional, and voiceprint-related features, which are often interdependent and contextually influenced \cite{pan2023msac,ju2024naturalspeech}—presents a persistent challenge. Consequently, the effective and comprehensive disentangling of these speech attributes into independent components remains a complex and ongoing task.

\section{Methodology}
\label{sec:METHODOLOGY}

\subsection{PSCodec-Base}
\subsubsection{System Overview}
Building upon the baseline HifiCodec \cite{yang2023hifi}, \emph{PSCodec-Base} encodes any given input speech \( X \) into a low-dimensional representation space \( Z \in \mathbb{R}^{d \times f / M} \) using the Enc, where \( M \) represents its striding factor. 
The Enc comprises a one-dimensional convolutional (Conv1d) layer with 512 channels and a kernel size of 7, followed by four convolutional blocks. Each block incorporates a down-sampling module that employs a strided convolution, wherein the kernel size is twice the corresponding stride, alongside a residual module that consists of two convolutional operations and a skip connection. After these blocks, a final Conv1d layer with a kernel size of 3 and 512 output channels is applied to capture the encoded features $Z$. 
Subsequently, a group residual vector quantization (GRVQ) based Quan is employed to generate the quantized features \( Z_q \in \mathbb{R}^{d \times f / M} \) via learnable codebook representations.
Simultaneously, two effective prompt encoders are introduced to extract the VP-related features \( Z_{vp} \) and Mel-related features \( Z_{mel} \) from the input speech prompt \( X_P \). 
Last, the reconstructed waveform \( \tilde{X} \) is obtained by performing element-wise summation of \( Z_q \), \( Z_{vp} \), and \( Z_{mel} \), followed by processing the combined representation through the Dec with an up-sampling rate of \( M \) as well.
Notably, the down-sampling strides of the Enc are configured as [2, 4, 5, 8], while the up-sampling strides of the Dec are [4, 4, 4, 5].

\subsubsection{Prompt Encoders}
To enhance the performance of neural speech codecs at low bitrates, we develop two specialized prompt encoders leveraging artificial statistical features. The intuition behind is that by first decoupling and subsequently integrating certain representations from the speech signal, the speech codec can effectively reduce the information volume to be processed, thereby enhancing both compression efficiency and overall performance.

\begin{figure}
\centering
	\includegraphics[height=5.5cm,width=!]{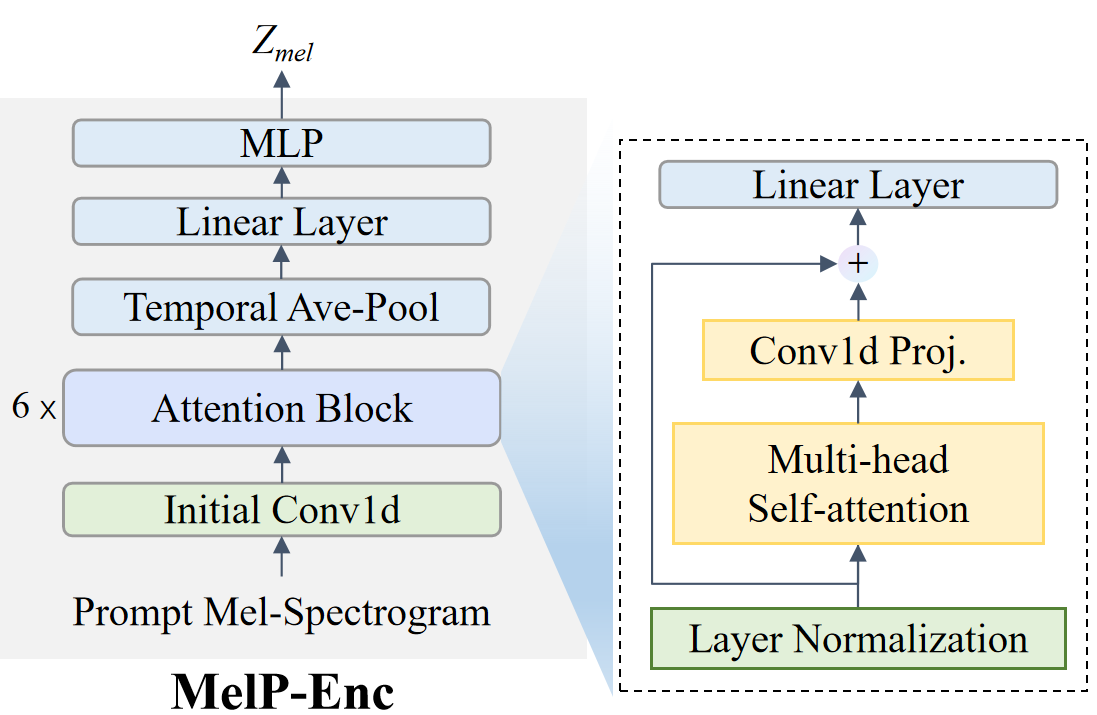}
	\caption{Detailed architecture of the proposed MelP-Enc.}
        \label{fig:cond_encoder}
\end{figure}

\textbf{MelP-Enc.}
Mel-related elements, including paralinguistic information, are fundamental components of human speech, encapsulating critical aspects such as emotional state, tone, and intonation. These features not only simulate human auditory perception through compressed representations but also provide valuable insights into the intricate nuances of speech. Consequently, we propose a learnable MelP-Enc to efficiently decouple and harness these features, with its detailed architecture depicted in Fig. \ref{fig:cond_encoder}.
To be precise, MelP-Enc initially employs a Conv1d layer to transform the Mels $X_{mel}$ of the input prompt $X_{P}$ into a latent feature space comprising 512 channels. 
Subsequently, to enhance the representational capacity of the hidden features, we design six attention-based blocks, each consisting of several components, including layer normalization (LN), multi-head self-attention (MHSA), Conv1d, linear projection, and skip connection operations. This design facilitates the effective integration of the extracted representations while capturing their intricate patterns.
Following these blocks, a temporal averaging operation is applied to fuse the captured features and then integrate a multi-layer projection (MLP) module to align its final representations $Z_{mel}$ with $Z_q$.
\begin{equation}
    \begin{split}
       Z_{mel} = MLP(Enc_{MelP}(X_{mel}))
    \end{split}
\end{equation}

\begin{figure*}[htbp]
\centering
	\includegraphics[height=6.9cm,width=!]{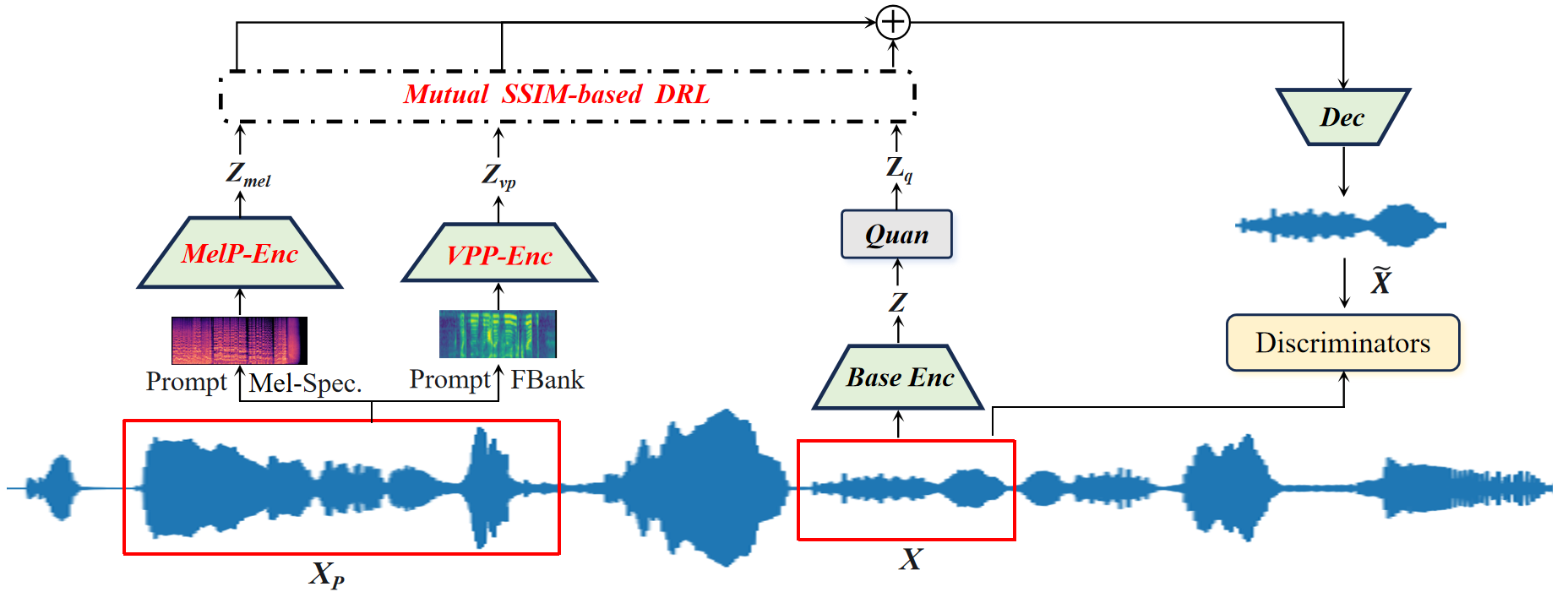}
        \caption{Overview of our proposed PSCodec-DRL-ICT training framework. The mutual SSIM-based DRL represents computing the SSIM scores between each pair of the three encoded features, i.e., $Z_{mel}$, $Z_{vp}$, and  $Z_{q}$.}        
        \label{fig:PSCodec-DRL-ICT}
\end{figure*}

\textbf{VPP-Enc.}
In addition to the Mel-associated information, VP features have traditionally been acknowledged as intrinsic and globally time-invariant components of speech signals.
Therefore, a natural and straightforward solution is to utilize a pre-trained SV model to extract and separate the VP features within utterances to further mitigate the load on the Base Enc.
In practice, we choose to employ a lightweight and efficient pre-trained SV model, CAM++ \cite{wang2023cam}.
Detailed, the FBank features $X_{F}$ of the input prompt speech $X_P$ are first fed into the CAM++ to extract the VP representations.
Afterwards, we likewise adopt a MLP module that consists of a stack of linear and activation layers to align $Z_{vp}$ with $Z_q$, as depicted in Fig. \ref{fig:PSCodec-base}. It is worth noting that all parameters of CAM++ are frozen throughout the training stage.
\begin{equation}
    \begin{split}
       Z_{vp} = MLP(Enc_{V\!PP}(X_F))
    \end{split}
\end{equation}

\subsection{PSCodec-DRL-ICT}

To improve the overall performance of PSCodec-Base, we introduce an SSIM-based DRL method alongside an ICT strategy to enhance its feature utilization efficiency. The detailed architecture is presented in Fig. \ref{fig:PSCodec-DRL-ICT}.

\subsubsection{SSIM-based DRL}

Under our proposed prompt encoders augmented workflow, using multiple encoders to decouple and harness specific speech attribute features inherently introduces redundancy among the encoding features. 
Therefore, to mitigate this issue and enhance its feature utilization efficiency, we propose a SSIM-based DRL method to optimize all the encoders (Enc, MelP-Enc, and VPP-Enc) and constrain their encoding representations.
Concretely, the SSIM metrics between each pair of $Z_q$, ${Z}_{vp}$ and ${Z}_{mel}$ are first calculated, which can be formulated as follows:
\begin{equation}
    \begin{gathered}
\text{SSIM}(Z_i, Z_j) = \frac{(2\mu_{Z_i} \mu_{Z_j} + c_1)(2\sigma_{{Z_i}{Z_j}} + c_2)}{(\mu_{Z_i}^2 + \mu_{Z_j}^2 + c_1)(\sigma_{Z_i}^2 + \sigma_{Z_j}^2 + c_2)}
    \end{gathered}
\end{equation}
where $\mu_{Z_i}$ and $\mu_{Z_j}$ represent the means of $Z_i$ and $Z_j$, $\sigma_{Z_i}^2$ and $\sigma_{Z_j}^2$ denote their variances, and $\sigma_{{Z_i}{Z_j}}$ refers to the covariance between $Z_i$ and $Z_j$. The constants $c_1$ and $c_2$ are introduced to stabilize the division, with their values set to $0.01$ and $0.03$, respectively.
Following this, we minimize their weighted sum as an additional penalty incorporated into the training loss, thereby encouraging each encoder to focus on predicting distinct information.
\begin{equation}
    \begin{gathered}
        L_1 = \text{SSIM}(Z_q, {Z}_{mel}) \\
        L_2 = \text{SSIM}(Z_q, {Z}_{vp}) \\
        L_3 = \text{SSIM}({Z}_{vp}, {Z}_{mel}) \\
        L_{DRL} = \lambda_1 L_1 + \lambda_2 L_2 + \lambda_3 L_3
    \end{gathered}
\end{equation}
where $\lambda_{1}$, $\lambda_{2}$, and $\lambda_{3}$ are weight coefficients to adjust $L_1$, $L_2$, and $L_3$. 
In our case, the values of $\lambda_{1}$, $\lambda_{2}$, and $\lambda_{3}$ are empirically set to 2, 5, and 1, respectively.

\subsubsection{Incremental Continuous Training (ICT)}
\label{subsec:Incremental Continuous Training Strategy}

While the integration of SSIM-based DRL effectively reduces redundancy between encoding features, training the proposed approach from scratch often leads to underfitting, especially in low-bitrate scenarios, where this issue is more pronounced. 

After a systematic analysis, we discover that directly optimizing all encoders with SSIM-based DRL from scratch tends to cause an overemphasis on reducing redundancy, hindering the model's ability to learn the essential information for speech reconstruction.
To address this problem, we present a two-stage ICT approach that ensures stable and efficient training, with its pseudo-code described in Algorithm \ref{alg:ict_training}. 
\begin{algorithm}[htbp]
\caption{ICT for \emph{PSCodec-DRL-ICT}}
\label{alg:ict_training}
\textbf{Input}: Training data $\mathcal{D}$, initial model parameters $\theta_0$\\
\textbf{Output}: Well-trained parameters of final stage  $\theta^*$

\begin{algorithmic}[1]
\STATE \textbf{Stage 1: \emph{PSCodec-Base Training}}
\begin{itemize}
\item Initialize PSCodec-Base with $\theta_0$
\item Train on $\mathcal{D}$, yielding parameters $\theta_1$
\end{itemize}
\STATE \textbf{Stage 2: \emph{PSCodec-DRL-ICT Training}}
\begin{itemize}
\item Initialize a new PSCodec-Base with $\theta_0$
\item Transfer Quan and all encoder parameters from $\theta_1$
\item Retrain using SSIM-based DRL on $\mathcal{D}$, yielding $\theta^*$
\end{itemize}
\STATE \textbf{return} $\theta^*$
\end{algorithmic}
\end{algorithm}
As shown in the figure, this process begins by training the PSCodec-Base model from scratch. In the second stage, the parameters of Quan and all encoders (including Enc, VPP-Enc, and MelP-Enc) from the well-trained PSCodec-Base in the first stage are transferred to a new PSCodec-Base model with the same architecture. This model is subsequently retrained using SSIM-based DRL to ensure a thorough and comprehensive update of all parameters.

\subsection{PSCodec-CasAN}

Although PSCodec-DRL-ICT exhibits great performance, its extensive hyperparameter tuning and multi-stage training make it somewhat labor-intensive. To facilitate a streamline training process, we introduce an advanced CasAN that effectively enables the adaptive fusion of encoded features, thereby enhancing overall system performance.

\begin{figure}[h]
\centering
	\includegraphics[height=11.9cm,width=!]{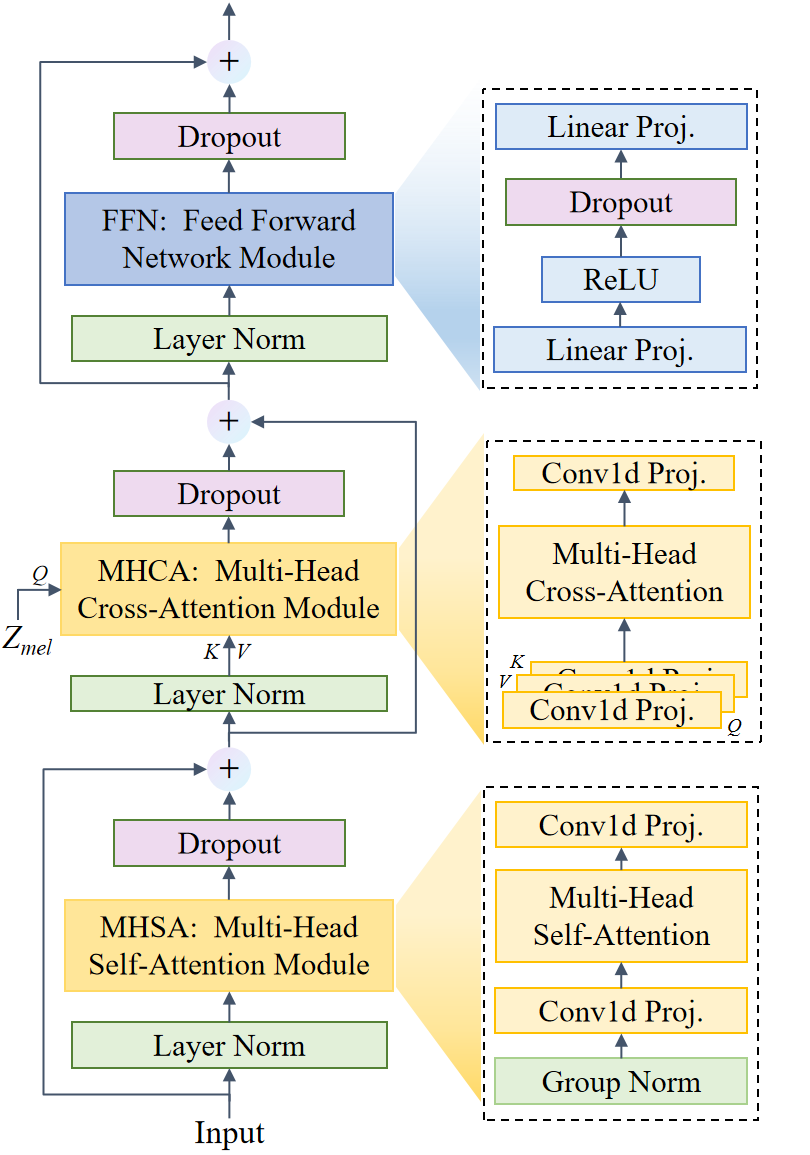}
	\caption{Detailed architecture of the proposed cascade attention module.}
        \label{fig:casan}
\end{figure}

To be specific, the proposed CasAN model comprises six cascaded attention blocks, each mainly consisting of three components: a MHSA module, a multi-head cross-attention (MHCA) module, and a feed-forward network (FFN) module. 
In addition to these components, LN is applied prior to each module, while dropout is employed afterward to facilitate regularization and enhance model generalization.

Considering that the voiceprint feature $Z_{vp}$ is typically regarded as a global time-invariant feature \cite{wang2023cam,pan2024ctefm}, we initially combine the quantized features $Z_q$ and $Z_{vp}$ through element-wise summation, yielding the feature $Z_{q-vp}$, which serves as input to the first cascaded attention block.
Subsequent cascaded attention blocks receive as input the outputs from their preceding block.
Next, we regularize $Z_{q-vp}$ with LN before feeding it into the MHSA module with dropout. This enables the adaptive fusion and capture of the relationship between $Z_q$ and $Z_{vp}$, thus enhancing the representational capacity of the $Z_{q-vp}$ feature.
As illustrated in Fig. \ref{fig:casan}, the MHSA module incorporates a group normalization layer, flanked by two Conv1d projection layers in a macaron-like structure, surrounding the conventional MHSA mechanism \cite{vaswani2017attention,yang2023hybridformer}.
The output features from the MHSA module, regularized by LN, are then passed into the MHCA module as key ($K$) and value ($V$), with $Z_{mel}$ serving as the query ($Q$) for the MHCA module:
\begin{equation}
    \begin{split}
       \emph{MHCA}(Q, K, V) = \emph{Concat}(\emph{head}_1, \dots, \emph{head}_h)
    \end{split}
\end{equation}
\begin{equation}
    \begin{split}
       \emph{head}_i = \text{Attention}(Q, K, V) = \emph{Softmax}\left(\frac{Q K^T}{\sqrt{d_k}}\right) V
    \end{split}
\end{equation}
where $d_k$ denotes the dimensionality of the $K$ vectors and $h$ represents the number of attention heads, with values set to 128 and 4, respectively, in our implementation.

Finally, the output features from the MHCA module are passed into the FFN module, which is composed of two linear layers, with a ReLU activation function and a dropout layer applied in between.

\subsection{Training Criteria}
The training objectives of our proposed PSCodec systems are composed of the generator loss and the discriminator loss.
Regarding the discriminator loss, a generative adversarial network (GAN) based training approach is adopted for all the proposed methods.
We empirically utilize MPD \cite{kong2020hifi} and MS-STFTD \cite{kumar2024high} to promote the perceptual quality of the reconstructed waveform:
\begin{equation}
    \begin{gathered}
         \mathcal{L}_{\text{ad}} = \frac{1}{K} \sum_{k=1}^K \max(0, 1 - D_k(X)) + \max(0, 1 + D_k(\tilde{X}))
    \end{gathered}
\end{equation}
where \( K \) denotes the total number of discriminators, with \( D_k \) representing the \( k \)-th discriminator. 

The generator training loss of PSCodec-Base and PSCodec-CasAN comprises three components which are the conventional reconstruction loss $L_{r}$, feature matching loss $L_{f}$, GRVQ commitment loss $L_{q}$.
In contrast, in addition to these three losses, the generator training loss for PSCodec-DRL-ICT includes an additional SSIM-based DRL loss $L_{DRL}$.
For the reconstruction loss $L_{r}$, we first compute the $l_1$-norm distance between the input speech waveform and the reconstructed waveform. Subsequently, we calculate the $l_1$-norm distance between the corresponding Mels, which are obtained by applying window lengths of [256, 512, 1024, 2048, 4096] and setting the hop length to one-quarter of the window length.
\begin{equation}
    \begin{gathered}
        L_{r} = \left\| X - \tilde{X} \right\|_1 + 
                  \sum_{i=1}^N \left\| \emph{Mels}(X) - \emph{Mels}(\tilde{X}) \right\|_1
    \end{gathered}
\end{equation}
where \emph{Mels($\cdot$)} represents the standard calculation function of Mel-spectrogram, and $N$ equals 5.
The $L_{f}$ is computed as the average distance between the \( l \)-th hidden features of the \( k \)-th discriminator:
\begin{equation}
    \begin{gathered}
         L_{\emph{f}} = \frac{1}{K L} \sum_{k} \sum_{l} \left\| D_k^l(X) - D_k^l(\tilde{X}) \right\|_1
    \end{gathered}
\end{equation}
The GRVQ commitment loss $L_{q}$ can be formulated as:
\begin{equation}
    \begin{gathered}
        L_q(Z, Z_q) = \sum_{i=1}^N \left\| Z_i - \hat{Z}_i \right\|_2^2
    \end{gathered}
\end{equation}

As a consequence, the final training loss of PSCodec-Base, PSCodec-CasAN, and PSCodec-DRL-ICT can be defined as:
\begin{equation}
    \begin{gathered}
        L_{Base/CasAN} = \!\beta_1 \!L_{r} + \!\beta_2 \!L_{f} + \!\beta_3 \!L_{q} + \!\beta_4 \!L_{ad}
    \end{gathered}
\end{equation}
\begin{equation}
    \begin{gathered}
        L_{DRL} = \!\beta_1 \!L_{r} + \!\beta_2 \!L_{f} + \!\beta_3 \!L_{q} + \!\beta_4 \!L_{ad} + \!\beta_5 \!L_{DRL}
    \end{gathered}
\end{equation}
where $\beta_{1}$, $\beta_{2}$, $\beta_{3}$, $\beta_{4}$, and $\beta_{5}$ are hyper-parameters, with the values of 2, 1, 50, 1, and 1, respectively, in our experiments.

\section{Experiments}
\label{sec:EXPERIMENTS}

\subsection{Experimental Setups}
\subsubsection{Dataset}
To ensure a fair comparison with SOTA neural speech codecs, all experiments are conducted on the LibriTTS dataset \cite{zen2019libritts}, which comprises 585 hours of recordings from 2,456 English speakers. We use a combination of the train-other-500, train-clean-360, and train-clean-100 subsets for training both the proposed PSCodec-based methods and the comparative models. Additionally, to provide a comprehensive evaluation of the performance of the proposed systems, we select the test-clean and test-other subsets of LibriTTS as ID test sets, along with the LJSpeech dataset as an OOD test set.

\begin{table*}[htbp]
    \centering
    \caption{ID performance comparison of SOTA neural codecs on the LibriTTS test-clean and test-other datasets. The \\optimal results are highlighted in bold, while the sub-optimal results are underlined.}
    \label{tab:in-domain-results-clean}
    \begin{tabular}{cccccccccc}
        \toprule
        Models    &   $N_q$    &   Bitrate    &   PESQ ($\uparrow$) & STOI ($\uparrow$) & ViSQOL ($\uparrow$) & UTMOS ($\uparrow$) & SiSNR ($\uparrow$) & MCD ($\downarrow$) & SECS ($\uparrow$) \\
        \midrule
        \multicolumn{9}{l}{LibriTTS Test-Clean} \\
        \midrule
        AudioDec  & \multirow{9}{*}{1}  & \multirow{9}{*}{675 bps}  & 1.168  &  0.294  &  1.161  &  1.395   &    -36.548  &  4.007  &  0.431  \\
        APCodec    &  &  &   1.201  &  0.747  &  1.426  &  2.476  &  -39.924  &  3.020  &  0.537  \\
        DAC    &  &  &       1.596  &  0.839  &  1.726  &  2.893   &    -1.791  &  1.852  &  0.732  \\
        TiCodec   &  &  &        1.882  &  0.878  &  1.920  &  3.755   &    -2.618  &  1.621  &  0.788  \\
        Encodec  &  &  &   2.030  &  0.910  &  2.774 &   2.706  & -27.471  &  1.361  &  0.715  \\
        HifiCodec &  &  &    2.359  &  0.914  &  2.352 &   3.731  &  0.776  &  1.262  &  0.864  \\
        PSCodec-Base   &  &   &  2.684  &  0.932  &  3.306 &  3.986  &  1.715  &  1.181  &  0.870  \\
        PSCodec-DRL-ICT   &  &   &    \textbf{2.786}  &  \textbf{0.942}  &  \underline{3.447}  &  \underline{4.039}  &   \textbf{2.526}  &  \textbf{1.086}  &  \underline{0.884}  \\
        PSCodec-CasAN   &  &   &    \underline{2.750}  &  \underline{0.941}  &  \textbf{3.452}  &   \textbf{4.048}  &   \underline{2.447}  &  \underline{1.093}  &  \textbf{0.888}  \\
        \midrule[\heavyrulewidth]
        AudioDec  & \multirow{9}{*}{2}  & \multirow{9}{*}{1.35 kbps} &    1.248  &  0.321  &  1.685  &   1.637  &    -34.104  &  3.924  &  0.524  \\
        APCodec    &  &  &   1.404  &  0.841  &  2.260  &  2.885  &  -20.609  &  2.268  &  0.690  \\
        DAC    &  &  &        2.342  &  0.917  &  2.632  &  3.739   &   2.746  &  1.233  &  0.847  \\
        TiCodec   &  &  &     2.405  &  0.921  &  2.521  &  3.872   &   1.179  &  1.235  &  0.858  \\
        Encodec  & & &    2.447  &  0.936  &  3.430 &  3.348   &    -7.345  &  1.131  &  0.790  \\
        HifiCodec &  &  &   3.024  &  0.949  &  3.088  &  3.853   &  3.571  &  0.931  &  0.900  \\
        PSCodec-Base   &  &   &  3.245  &  0.962  &  3.807 &  4.002  &  3.496  &  0.937  &  0.906  \\
        PSCodec-DRL-ICT   &  &   &  \textbf{3.398}  &  \textbf{0.966}  &  \textbf{3.954}  &  \underline{4.063}   &   \textbf{4.829}  &  \textbf{0.823}  &  \textbf{0.935}  \\
        PSCodec-CasAN   &  &   &  \underline{3.391}  &  \textbf{0.966}  &  \underline{3.837}  &  \textbf{4.067}   &   \underline{4.294}  &  \underline{0.827}  &  \underline{0.921}  \\
        \midrule
        \multicolumn{9}{l}{LibriTTS Test-Other} \\
        \midrule
        AudioDec  & \multirow{9}{*}{1}  & \multirow{9}{*}{675 bps} &    1.150  &  0.648  &  1.425  &   1.714  &    -33.573  &  3.231  &  0.516  \\
        APCodec    &  &  &   1.197  &  0.717  &  1.353  &  2.271  &  -38.357  &  3.409  &  0.422  \\
        DAC    &  &  &   1.502  &  0.808  &  1.672  &  2.511   &   -1.372 &  1.974  &  0.661   \\
        TiCodec  &  &  &    1.685  &  0.837  &  1.811  &   3.141  &   -2.683 &  1.840  &  0.727   \\
        Encodec  &  &  &  1.922  &  0.885  &  2.584 &  2.349  &    -25.542  &  1.464  &  0.653  \\
        HifiCodec &  &  &    2.083  &  0.881  &  2.211 &   3.107  &  0.594  &  1.426  &  0.807  \\
        PSCodec-Base   &  &   &2.426  &  0.914  &  3.087 &  3.467  &  1.736  &  1.309  &  0.825   \\
        PSCodec-DRL-ICT   &  &   &   \textbf{2.497}  &  \textbf{0.919}  &  \textbf{3.335}  &  \underline{3.517}   &   \textbf{2.506}  &  \textbf{1.234}  &  \underline{0.838}  \\
        PSCodec-CasAN   &  &   &   \underline{2.449}  &  \underline{0.916}  &  \underline{3.292}  &  \textbf{3.544}   &   \underline{2.290}  &  \underline{1.253}  &  \textbf{0.844}  \\
        \midrule[\heavyrulewidth]
        AudioDec  & \multirow{9}{*}{2}  & \multirow{9}{*}{1.35 kbps} &    1.338  &  0.702  &  1.813  &   2.171  &   -30.068  &  2.783  &  0.649   \\
        APCodec    &  &  &   1.348  &  0.806  &  2.142  &  2.476  &  -22.870  &  2.794  &  0.612  \\
        DAC   &  &  &     2.112  &  0.891  &  2.495  &  3.206  &  3.075  &  1.344  &  0.794   \\
        TiCodec  &  &  &     2.122  &  0.888  &  2.347  &   3.269  &   1.015  &  1.403  &  0.803   \\
        Encodec  &  &  &  2.268  &  0.914  &  3.265 &  2.933  &    -7.429  &  1.218  &  0.735  \\
        HifiCodec &  &  &   2.687  &  0.926  &  3.000  &  3.234   &  3.419  &  1.149  &  0.852  \\
        PSCodec-Base   &  &   &  2.882  &  0.935  &  3.661 & 3.456  &  3.472  &  1.079  &  0.869  \\
        PSCodec-DRL-ICT   &  &   &  \textbf{3.063}  &  \textbf{0.949}  &  \textbf{3.911}  &   \underline{3.530}  &   \textbf{4.666}  &  \underline{0.951}  &  \textbf{0.901}  \\
        PSCodec-CasAN   &  &   &  \underline{3.055}  &  \textbf{0.949}  &  \underline{3.776}  &  \textbf{3.550}   &   \underline{4.248}  &  \textbf{0.949}  &  \underline{0.885}  \\
        \bottomrule
    \end{tabular}
\end{table*}

\subsubsection{Implementation Details}

In all experiments, we use the Adam optimizer \cite{kingma2014adam} with an initial learning rate of 1e-3 and a batch size of 40 to train the proposed three PSCodec-based systems and other codecs on two A10 GPUs for 500K steps.
The learning rate is scheduled using the OneCycleLR policy, with a maximum momentum of 0.98.
For the model configuration, we set the decoder dimension and codebook size of all approaches to 512. In total, our PSCodec-Base, PSCodec-DRL-ICT, and PSCodec-DRL-ICT have 84M, 84M, and 109M parameters, respectively.
During training, all data, originally sampled at 24 kHz, is randomly segmented within each batch into two speech snippets: a 1-second input speech segment and a 3-second input speech prompt.
For the input speech prompt, its corresponding Mel and Fbank features are extracted. The input prompts used for extracting Fbank features are subsequently resampled to a 16kHz sampling rate due to the employed CAM++ model\footnote{\url{https://modelscope.cn/models/iic/speech_campplus_sv_zh_en_16k-common_advanced}}, which is accessible on an open-source website.
It is worth noting that during inference, to simulate real-world application scenarios, the input speech prompt is a different utterance from the same speaker as the input speech.

\subsubsection{Baselines}
To thoroughly assess the effectiveness of the proposed PSCodec systems, we adopt 6 SOTA neural speech codec methods, i.e., Encodec, HifiCodec, TiCodec, DAC, AudioDec, and APCodec as the baselines.
For all baselines, we re-implement them using the open-source implementations\footnote{\url{https://github.com/yangdongchao/AcademiCodec}}\footnote{\url{https://github.com/y-ren16/TiCodec}}\footnote{\url{https://github.com/descriptinc/descript-audio-codec}}\footnote{\url{https://github.com/facebookresearch/AudioDec}}\footnote{\url{https://github.com/YangAi520/APCodec}}, retaining all of their original configurations except for the decoder dimension and codebook size for fair comparisons.

\subsubsection{Evaluation Metrics}
To make an in-depth evaluation, we employ multiple metrics to verify the speech reconstruction and speaker similarity performance of all neural speech codecs, including the ViSQOL, PESQ, STOI, SiSNR, UTMOS, MCD, and SECS.
Concretely, the STOI aims to test the intelligibility of the reconstructed speech in comparison to the initial utterance. 
PESQ, ViSQOL, and UTMOS are used to examine the overall perceived quality of neural codecs, while SiSDR and MCD are adopted to assess the phase and amplitude spectrum quality, respectively.
Furthermore, we also utilize the pre-trained CAM++ to measure the SECS to quantify the speaker similarity between the original input and generated speech, which closely aligns with the subjective perceptions of auditory evaluators.

\begin{table*}[htbp]
    \centering
    \caption{Overall OOD performance comparison of SOTA neural speech codecs on the LJSpeech dataset. The bolded numbers \\represent the optimal results, while the underlined numbers indicate the sub-optimal results.}
    \label{tab:out-of-domain-results}
    \begin{tabular}{cccccccccc}
        \toprule
        Models    &   $N_q$    &   Bandwidth    &   PESQ ($\uparrow$) & STOI ($\uparrow$) & ViSQOL ($\uparrow$) & UTMOS ($\uparrow$) & SiSNR ($\uparrow$) & MCD ($\downarrow$) & SECS ($\uparrow$) \\
        \midrule
        AudioDec  & \multirow{9}{*}{1}  & \multirow{9}{*}{675 bps} &    1.166  &  0.668  &  1.270  & 1.878 &  -35.746  &  3.231  &  0.586   \\
        APCodec    &  &  &   1.163  &  0.723  &  1.115  &  2.451  &  -40.803  &  3.336  &  0.568  \\
        DAC   &  &  &  1.633  &  0.853  &  1.640  &  3.112  &   -1.194  &  1.956  &  0.816  \\
        TiCodec  &   &   &    1.824  &  0.877  &  1.675  &  3.784 &  -3.635  &  1.822  &  0.814   \\
        Encodec  &  &  &  2.022  &  0.917  &  2.702 &  2.873  &   -28.093  &  1.377  &  0.724  \\
        HifiCodec &  &  &    2.036  &  0.908  &  2.326 &  3.797  &  -0.976  &  1.466  &  0.881  \\
        PSCodec-Base   &  &   &    2.589  &  0.939  &  3.216 & 4.030  &  -0.636  &  1.276  &  0.915  \\
        PSCodec-DRL-ICT   &  &   &   \textbf{2.673}  &  \textbf{0.943}  &  \textbf{3.332}  & \textbf{4.333}  &   \textbf{-0.198}  &  \textbf{1.160}  &  \underline{0.925}  \\
        PSCodec-CasAN   &  &   &     \underline{2.585}  &  \underline{0.939}  &  \underline{3.267} & \underline{4.274}  &  \underline{-0.412}  &  \underline{1.211}  &  \textbf{0.926}  \\
        \midrule[\heavyrulewidth]
        AudioDec  & \multirow{9}{*}{2}  & \multirow{9}{*}{1.35 kbps} &    1.415  &  0.729  &  1.710  & 2.655 &  -32.146  &  2.708  &  0.746   \\
        APCodec    &  &  &   1.497  &  0.814  &  1.808  &  2.602  &  -26.656  &  2.801  &  0.715  \\
        DAC   &  &  &  2.413  &  0.925  &  2.618  &  4.185  &   2.700  &  1.319  &  0.889  \\
        TiCodec  &   &   &    2.188  &  0.911  &  2.256  &  3.998  &   -2.390  &  1.461  &  0.889   \\
        Encodec  &   &   &  2.413  &  0.938  &  3.339 &  3.597  &   -14.434  &  1.157  &  0.818  \\
        HifiCodec &  &  &   2.463  &  0.939  &  3.048  &  3.898  &  0.627  &  1.171  &  0.916  \\
        PSCodec-Base &  &  &  2.816  &  0.948  &  3.752  &  4.217  &  0.737  &  1.094  &  0.939  \\
        PSCodec-DRL-ICT   &  &   &  \textbf{3.067}  &  \textbf{0.962}  &  \textbf{3.942}  &  \textbf{4.340} &   \textbf{2.150}  &  \textbf{0.935}  &  \textbf{0.961}  \\
        PSCodec-CasAN &  &  &  \underline{2.984}  &  \underline{0.959}  &  \underline{3.797}  &  \underline{4.295}  &  \underline{1.290}  &  \underline{0.977}  &  \underline{0.956}  \\
        \midrule[\heavyrulewidth]
    \end{tabular}
\end{table*}

\subsection{Main Results}

We conduct a comparative analysis between the proposed PSCodec-based methods and SOTA neural speech codecs, with a primary focus on two crucial aspects: their comprehensive ID performance and OOD generalization capabilities.

\subsubsection{ID Evaluation.}

Table \ref{tab:in-domain-results-clean} compares the proposed PSCodec systems with SOTA neural codecs on the ID LibriTTS test-clean and test-other sets. 
The former contains 4,837 speech samples, while the latter, which features recordings in noisy environments, includes 5,120 samples. 
As presented in the tables, we can observe the following conclusions: 
\begin{enumerate}
\item It is evident that on both ID testing conditions, in contrast to SOTA neural codecs, our proposed PSCodec-based approaches consistently achieve the best results in all metrics assessing speech reconstruction and speaker similarity at various low bitrates. 
To be specific, PSCodec-DRL-ICT achieves the highest performance, with PSCodec-CasAN yielding comparable results and PSCodec-Base performing slightly lower than both. Among the baselines, HifiCodec demonstrates the best results, though it remains significantly below that of our proposed PSCodec-Base. Moreover, Encodec outperforms TiCodec and DAC, while AudioDec exhibits a slightly lower performance compared to APcodec.
\item At a bitrate of 1350 bps with 2 codebooks, the proposed three PSCodec methods simultaneously outperforms other codecs across all metrics on both ID test sets, underscoring their superior capacities.
Concretely, on the LibriTTS test-clean set, the best performing PSCodec-DRL-ICT exhibits a 13.1\% relative reduction in MCD and relative improvements of 12.4\%, 1.8\%, 5.5\%, 28.0\%, 35.2\%, and 3.9\% in terms of PESQ, STOI, UTMOS, ViSQOL, SiSNR, and SECS, compared to the best baseline HifiCodec.
\item At a bitrate of 675 bps with one codebook, our three PSCodec-based systems consistently surpasses all neural codecs across various metrics on both ID test sets as well.
To elaborate, on the test-clean set, PSCodec-DRL-ICT outperforms HifiCodec by 18.1\%, 3.1\%, 8.3\%, 46.6\%, 225.5\%, and 2.3\% in PESQ, STOI, UTMOS, ViSQOL, SiSNR, and SECS, showing greater performance improvements as opposed to the scenario of 1350 bps bitrate.
\item We notice that the advantage of the proposed PSCodec systems becomes even more pronounced on the noisy test-other set under both 675 bps and 1350 bps bitrates, exhibiting better performance improvements in all metrics relative to the test-clean set. 
These findings validate the effectiveness and robustness of our proposed PSCodec-based approaches.
\end{enumerate}

\subsubsection{OOD Generalization Evaluation.}

Then, we assess the OOD generalization performance of all codecs using 13,100 utterances from the LJSpeech dataset, which are resampled from a native 22,050Hz to 24,000Hz during inference. All results are summarized in Table \ref{tab:out-of-domain-results}.
From the table, it is easily to draw the following findings:
\begin{enumerate}
\item Table \ref{tab:out-of-domain-results} provides clear evidence that among all codecs, the proposed PSCodec-based systems achieve the best results across all metrics, indicating their effectiveness and robustness. Furthermore, HifiCodec and Encodec exhibit superior performance compared to other baselines, whereas AudioDec and APCodec demonstrate relatively weaker outcomes. At a low bitrate of 675 bps, TiCodec surpasses DAC in performance; however, at a relatively higher bitrate of 1350 bps, DAC outperforms TiCodec.
\item As the bitrate decreases, the advantage of our PSCodec-based approaches also become increasingly pronounced, highlighting the superiority of the proposed efficient prompt encoder augmentation strategy.
\item All neural speech codecs produce consistent results and trends under both OOD and ID test sets, underscoring their robustness. Importantly, our proposed PSCodec-based approaches consistently achieve the best performance across all metrics under different testing scenarios with various low bandwidths, highlighting their exceptional capabilities.
\end{enumerate}

Overall, these evaluations consistently demonstrate that our proposed three PSCodec-based systems, equipped with efficiently integrated prompt encoders, deliver exceptional performance in speech reconstruction and speaker similarity across various testing conditions under low bitrate scenarios. Compared to SOTA neural speech codecs, these systems present a promising solution for real-world applications in speech compression, synthesis, and related tasks.

\subsection{Ablation Study}

\begin{table*}[htbp]
    \centering
    \caption{Ablation study of the proposed PSCodec-CasAN on the LibriTTS test-clean set.}
    \label{tab:ablation-casan}
    \begin{tabular}{cccccccccc}
        \toprule
        Models    &   $N_q$    &   Bandwidth    &   PESQ ($\uparrow$) & STOI ($\uparrow$) & ViSQOL ($\uparrow$) & UTMOS ($\uparrow$) & SiSNR ($\uparrow$) & MCD ($\downarrow$) & SECS ($\uparrow$) \\
        \midrule
        PSCodec-CasAN  & \multirow{2}{*}{1}  & \multirow{2}{*}{675 bps}    &  \textbf{2.750}  &  \textbf{0.941}  &  \textbf{3.452}  &   \textbf{4.048}  &   \textbf{2.447}  &  \textbf{1.093}  &  \textbf{0.888}  \\
        w/o CasAN    &  &  &    2.684  &  0.932  &  3.306 &  3.986  &  1.715  &  1.181  &  0.870  \\
        \midrule
        PSCodec-CasAN  & \multirow{2}{*}{2}  & \multirow{2}{*}{1.35 kbps}  & \textbf{3.391}  &  \textbf{0.966}  &  \textbf{3.837}  &  \textbf{4.067}   &  \textbf{4.294}  &  \textbf{0.827}  &  \textbf{0.921}  \\
        w/o CasAN   &  &  &    3.245  &  0.962  &  3.807 &  4.002  &  3.496  &  0.937  &  0.906  \\
        \bottomrule
    \end{tabular}
\end{table*}

\begin{table*}[htbp]
    \centering
    \caption{Ablation study of the proposed PSCodec-DRL-ICT on the LibriTTS test-clean set.}
    \label{tab:ablation-results-drlict}
    \begin{tabular}{cccccccccc}
        \toprule
        Models    &   $N_q$    &   Bandwidth    &   PESQ ($\uparrow$) & STOI ($\uparrow$) & ViSQOL ($\uparrow$) & UTMOS ($\uparrow$) & SiSNR ($\uparrow$) & MCD ($\downarrow$) & SECS ($\uparrow$) \\
        \midrule
        PSCodec-DRL-ICT  & \multirow{3}{*}{1}  & \multirow{3}{*}{675 bps}    &   \textbf{2.786}  &  \textbf{0.942}  &  \textbf{3.447}  & 4.039 &   \textbf{2.526}  &  \textbf{1.086}  &  \textbf{0.884}  \\
         w/o ICT     &  &   &    2.496  &  0.926  &  3.075  &  3.972 &  0.675  &  1.323  &  0.830  \\
        w/o DRL   &  &  &    2.704  &  0.938  &  3.406 & 3.993  &  1.735  &  1.151  &  0.870  \\
        \midrule
        PSCodec-DRL-ICT  & \multirow{3}{*}{2}  & \multirow{3}{*}{1.35 kbps}    & \textbf{3.398}  &  \textbf{0.966}  &  \textbf{3.954}  &   4.063  &   \textbf{4.829}  &  \textbf{0.823}  &  \textbf{0.935}  \\
         w/o ICT       &  &   &  3.220  &  0.957  &  3.727  &  3.998   & 3.632  &  0.950  &  0.906  \\
        w/o DRL    &  &  &    3.275  &  0.963  &  3.820 &   4.005  & 3.536  &  0.937  &  0.911  \\
        \bottomrule
    \end{tabular}
\end{table*}

To comprehensively evaluate the contributions of various components in our PSCodec-based systems, we conduct ablation studies.

\subsubsection{PSCodec-CasAN}

Table \ref{tab:ablation-casan} presents the results of an ablation study assessing the impact of the CasAN module within PSCodec-CasAN across different bitrates (675 bps and 1.35 kbps) on the LibriTTS test-clean set. The results clearly demonstrate that excluding CasAN consistently leads to performance degradation across all evaluation metrics. In detail, at a bitrate of 675 bps, removing CasAN results in performance declines ranging from $1.0\%$ to $29.9\%$, while at 1.35 kbps, the absence of CasAN similarly reduces performance by $0.4\%$ to $18.6\%$. 
Notably, SiSNR and MCD exhibit the most substantial performance decreases, underscoring the effectiveness of CasAN in enhancing the fidelity of the reconstructed speech. 
In a nutshell, these findings prove the essential role of CasAN in enhancing the codec's capacity to effectively integrate and utilize encoded features, thereby achieving substantial improvements in overall performance.

\subsubsection{PSCodec-DRL-ICT}

Regarding the proposed PSCodec-DRL-ICT, we first investigate the effectiveness of the proposed ICT approach.
As can be observed from Table \ref{tab:ablation-results-drlict}, omitting ICT results in a significant performance degradation at a bitrate of 675 bps, with reductions ranging from 1.7\% to 73.3\%. 
On the contrary, at a bitrate of 1.35 kbps, the exclusion of ICT leads to a more modest decline across all metrics, with performance reductions ranging from 0.9\% to 24.8\%. 
Hence, these results can prove the effectiveness of the ICT approach within the proposed PSCodec-DRL-ICT framework, particularly in more critical low-bitrate scenarios.

\begin{figure}[h]
\centering
	\includegraphics[height=5.7cm,width=!]{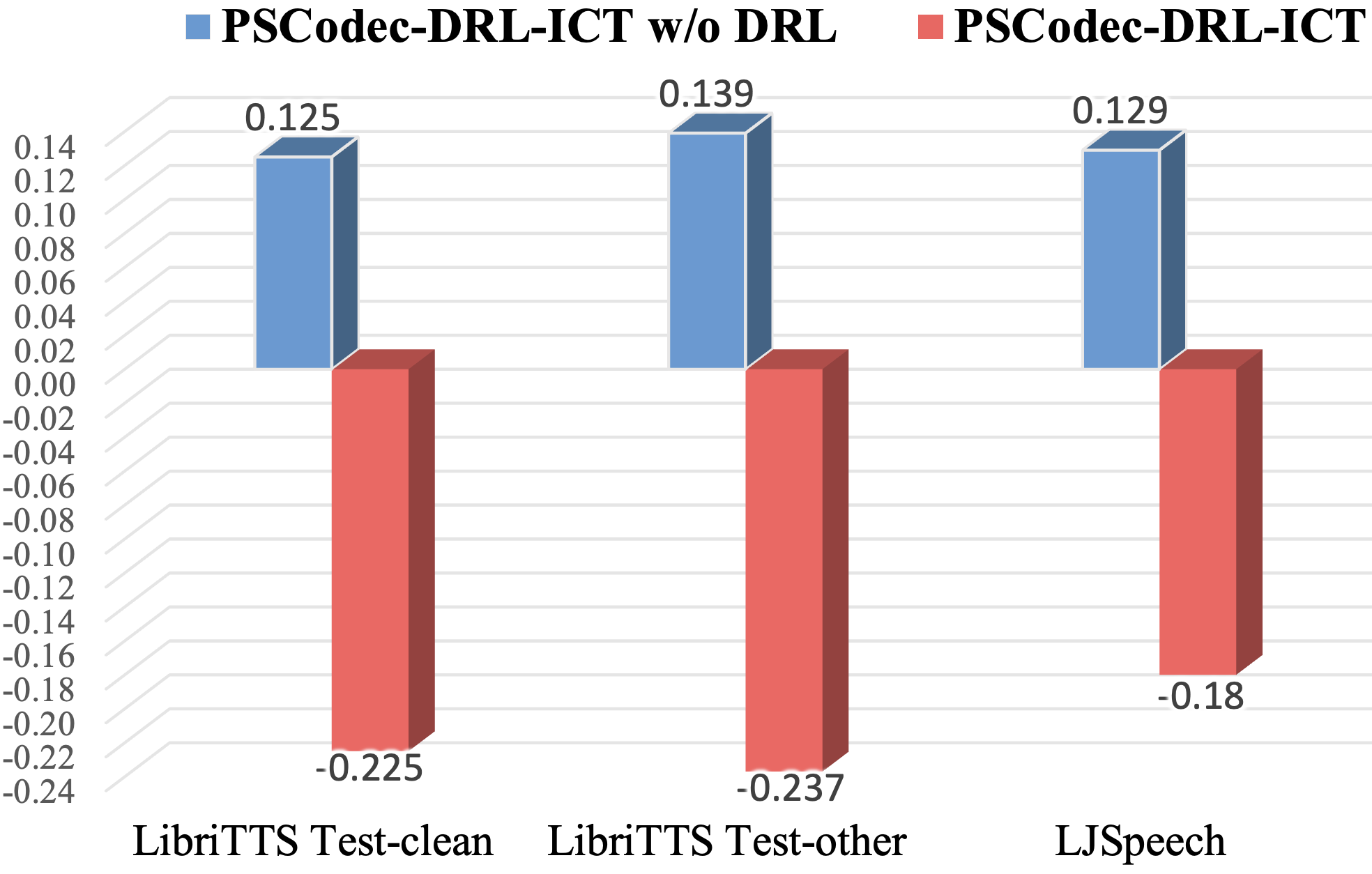}
	\caption{Comparison of Feature Utilization Efficiency Between PSCodec-DRL-ICT and Its Variant at a 675 bps Bitrate.}
        \label{fig:feature}
\end{figure}

\begin{table*}[htbp]
    \centering
    \caption{Ablation study of the proposed PSCodec-Base on the LibriTTS test-clean set.}
    \label{tab:ablation-results-base}
    \begin{tabular}{cccccccccc}
        \toprule
        Models    &   $N_q$    &   Bandwidth    &   PESQ ($\uparrow$) & STOI ($\uparrow$) & ViSQOL ($\uparrow$) & UTMOS ($\uparrow$) & SiSNR ($\uparrow$) & MCD ($\downarrow$) & SECS ($\uparrow$) \\
        \midrule
        PSCodec-Base   & \multirow{3}{*}{1}  & \multirow{3}{*}{675 bps}  &    \textbf{2.684}  &  \textbf{0.932}  &  \textbf{3.306} &  \textbf{3.986}  &  \textbf{1.715}  &  \textbf{1.181}  &  \textbf{0.870}  \\
        w/o VPP-Enc   &  &   &  2.587  &  \textbf{0.932}  &  3.184   &  3.906  &  -0.134  &  1.279  &  0.849   \\
        w/o MelP-Enc  &  &   &  2.547  &  0.931  &  3.129  &  3.841  &  -0.037  &  1.279  &  0.855  \\
        \midrule
        PSCodec-Base   & \multirow{3}{*}{2}  & \multirow{3}{*}{1.35 kbps}  & \textbf{3.245}  &  \textbf{0.962}  &  \textbf{3.807} &  \textbf{4.002}  &  \textbf{3.496}  &  \textbf{0.937}  &  \textbf{0.906}  \\
        w/o VPP-Enc   &  &   &  3.197  &  0.960  &  3.551  &  3.995  &  2.928  & 0.982  &  0.871   \\
        w/o MelP-Enc  &  &   &  3.174  &  0.958  &  3.470  &  3.936  &  1.654  &  1.005  &  0.879  \\
        \bottomrule
    \end{tabular}
\end{table*}

Then, we assess the impact of the proposed SSIM-based DRL approach. 
Fig. \ref{fig:feature} provides a quantitative feature utilization efficiency comparison of PSCodec-DRL-ICT and its variant without the SSIM-based DRL. As illustrated in the figure, we can clearly observe that the average SSIM similarity of encoded features obtained by PSCodec-DRL-ICT is significantly lower than that of its variant, showing that our SSIM-based DRL effectively reduces the redundancy among encoded features.
In addition, the results in Table \ref{tab:ablation-results-drlict} also demonstrate a consistent decline in performance across all metrics without the SSIM-based DRL, further underscoring the significance of our SSIM-based DRL method and reinforcing its critical role in enhancing overall speech quality.

\subsubsection{PSCodec-Base}

Finally, for the PSCodec-Base method, we evaluate the impact of the proposed two prompt encoders, namely VPP-Enc and MelP-Enc, with the results summarized in Table \ref{tab:ablation-results-base}.

As can be seen from the table, without either MelP-Enc or VPP-Enc, the performance of PSCodec-Base drops significantly, validating that the proposed prompt encoder augmented strategy can effectively improve the holistic capacity of neural codec models. 
Besides, at both 675 bps and 1350 bps bitrates, we can easily notice that MelP-Enc performs better than VPP-Enc in terms of perceptual evaluation metrics, while showing slightly inferior performance in speaker similarity. 
This discrepancy can likely be attributed to the fact that the prompt encoder must decouple and provide the decoder with supplementary information. Since the feature extractor of VPP-Enc (CAM++) is frozen, its efficacy is reduced in comparison to the more adaptive MelP-Enc. However, in terms of speaker similarity, VPP-Enc retains its advantage, likely due to its ability to capture speaker-specific features more effectively, as the frozen CAM++ module ensures more stable speaker identity representation during encoding.

\section{Conclusions}
\label{sec:CONCLUSIONS}
In this study, we present a series of scalable neural speech codec frameworks—PSCodec-Base, PSCodec-DRL-ICT, and PSCodec-CasAN—leveraging the efficient integration of effective prompt encoders to deliver high-performance speech reconstruction with low bandwidths. 
Specifically, we first propose PSCodec-Base that incorporates specialized prompt encoders to effectively disentangle and integrate the voiceprint and Mel-spectrogram-related representations, thereby improving both compression efficiency and speech reconstruction performance. 
Building on this, PSCodec-DRL-ICT introduces a SSIM-based DRL along with an ICT approach to enhance feature utilization and model robustness. 
Furthermore, to facilitate a more streamlined and efficient training process, PSCodec-CasAN incorporates an advanced cascaded attention network, further enhancing the representational capacity of the entire system.

Through extensive experimentation, we demonstrate that all of our proposed PSCodec frameworks outperform existing SOTA neural codec methods across various evaluation metrics, including speech quality, intelligibility, and speaker similarity. The results from our ablation studies and comparative analyses underscore the pivotal role of each proposed component in achieving the observed performance gains. Importantly, the advantages of the proposed PSCodec frameworks become more pronounced at lower bitrates compared to other SOTA codecs, highlighting their potential as highly effective solutions for practical applications in speech compression and generation.

\bibliographystyle{ieee}
\bibliography{pscodec}

\begin{thebibliography}{10}
\providecommand{\url}[1]{#1}
\csname url@samestyle\endcsname
\providecommand{\newblock}{\relax}
\providecommand{\bibinfo}[2]{#2}
\providecommand{\BIBentrySTDinterwordspacing}{\spaceskip=0pt\relax}
\providecommand{\BIBentryALTinterwordstretchfactor}{4}
\providecommand{\BIBentryALTinterwordspacing}{\spaceskip=\fontdimen2\font plus
\BIBentryALTinterwordstretchfactor\fontdimen3\font minus \fontdimen4\font\relax}
\providecommand{\BIBforeignlanguage}[2]{{%
\expandafter\ifx\csname l@#1\endcsname\relax
\typeout{** WARNING: IEEEtran.bst: No hyphenation pattern has been}%
\typeout{** loaded for the language `#1'. Using the pattern for}%
\typeout{** the default language instead.}%
\else
\language=\csname l@#1\endcsname
\fi
#2}}
\providecommand{\BIBdecl}{\relax}
\BIBdecl

\bibitem{xue2024low}
H.~Xue, X.~Peng, and Y.~Lu, ``Low-latency speech enhancement via speech token generation,'' in \emph{ICASSP 2024-2024 IEEE International Conference on Acoustics, Speech and Signal Processing (ICASSP)}.\hskip 1em plus 0.5em minus 0.4em\relax IEEE, 2024, pp. 661--665.

\bibitem{chiang2024restorative}
H.-T. Chiang, H.~Zhang, Y.~Xu\emph{,~et~al.}, ``Restorative speech enhancement: A progressive approach using se and codec modules,'' \emph{arXiv preprint arXiv:2410.01150}, 2024.

\bibitem{buthe2024nolace}
J.~B{\"u}the, A.~Mustafa, J.-M. Valin\emph{,~et~al.}, ``Nolace: Improving low-complexity speech codec enhancement through adaptive temporal shaping,'' in \emph{ICASSP 2024-2024 IEEE International Conference on Acoustics, Speech and Signal Processing (ICASSP)}.\hskip 1em plus 0.5em minus 0.4em\relax IEEE, 2024, pp. 476--480.

\bibitem{wang2023neural}
C.~Wang, S.~Chen, Y.~Wu\emph{,~et~al.}, ``Neural codec language models are zero-shot text to speech synthesizers,'' \emph{arXiv preprint arXiv:2301.02111}, 2023.

\bibitem{betker2023better}
J.~Betker, ``Better speech synthesis through scaling,'' \emph{arXiv preprint arXiv:2305.07243}, 2023.

\bibitem{borsos2023audiolm}
Z.~Borsos, R.~Marinier, D.~Vincent\emph{,~et~al.}, ``Audiolm: a language modeling approach to audio generation,'' \emph{IEEE/ACM Transactions on Audio, Speech, and Language Processing}, 2023.

\bibitem{zhang2023speak}
Z.~Zhang, L.~Zhou, C.~Wang\emph{,~et~al.}, ``Speak foreign languages with your own voice: Cross-lingual neural codec language modeling,'' \emph{arXiv preprint arXiv:2303.03926}, 2023.

\bibitem{wang2023lm}
Z.~Wang, Y.~Chen, L.~Xie\emph{,~et~al.}, ``Lm-vc: Zero-shot voice conversion via speech generation based on language models,'' \emph{IEEE Signal Processing Letters}, 2023.

\bibitem{martin2024enhancing}
{\'A}.~Mart{\'\i}n-Cortinas, D.~S{\'a}ez-Trigueros, I.~Vall{\'e}s-P{\'e}rez\emph{,~et~al.}, ``Enhancing the stability of llm-based speech generation systems through self-supervised representations,'' \emph{arXiv preprint arXiv:2402.03407}, 2024.

\bibitem{yang2024takin}
Y.~Yang, Y.~Pan, J.~Yao\emph{,~et~al.}, ``Takin-vc: Zero-shot voice conversion via jointly hybrid content and memory-augmented context-aware timbre modeling,'' \emph{arXiv preprint arXiv:2410.01350}, 2024.

\bibitem{chen2024takin}
S.~Chen, Y.~Feng, L.~He\emph{,~et~al.}, ``Takin: A cohort of superior quality zero-shot speech generation models,'' \emph{arXiv preprint arXiv:2409.12139}, 2024.

\bibitem{garbacea2019low}
C.~G{\^a}rbacea, A.~van~den Oord, Y.~Li\emph{,~et~al.}, ``Low bit-rate speech coding with vq-vae and a wavenet decoder,'' in \emph{ICASSP 2019-2019 IEEE International Conference on Acoustics, Speech and Signal Processing (ICASSP)}.\hskip 1em plus 0.5em minus 0.4em\relax IEEE, 2019, pp. 735--739.

\bibitem{zeghidour2021soundstream}
N.~Zeghidour, A.~Luebs, A.~Omran\emph{,~et~al.}, ``Soundstream: An end-to-end neural audio codec,'' \emph{IEEE/ACM Transactions on Audio, Speech, and Language Processing}, vol.~30, pp. 495--507, 2021.

\bibitem{defossez2022high}
A.~D{\'e}fossez, J.~Copet, G.~Synnaeve\emph{,~et~al.}, ``High fidelity neural audio compression,'' \emph{arXiv preprint arXiv:2210.13438}, 2022.

\bibitem{yang2023hifi}
D.~Yang, S.~Liu, R.~Huang\emph{,~et~al.}, ``Hifi-codec: Group-residual vector quantization for high fidelity audio codec,'' \emph{arXiv preprint arXiv:2305.02765}, 2023.

\bibitem{jenrungrot2023lmcodec}
T.~Jenrungrot, M.~Chinen, W.~B. Kleijn\emph{,~et~al.}, ``Lmcodec: A low bitrate speech codec with causal transformer models,'' in \emph{ICASSP 2023-2023 IEEE International Conference on Acoustics, Speech and Signal Processing (ICASSP)}.\hskip 1em plus 0.5em minus 0.4em\relax IEEE, 2023, pp. 1--5.

\bibitem{du2024funcodec}
Z.~Du, S.~Zhang, K.~Hu\emph{,~et~al.}, ``Funcodec: A fundamental, reproducible and integrable open-source toolkit for neural speech codec,'' in \emph{ICASSP 2024-2024 IEEE International Conference on Acoustics, Speech and Signal Processing (ICASSP)}.\hskip 1em plus 0.5em minus 0.4em\relax IEEE, 2024, pp. 591--595.

\bibitem{dietz2015overview}
M.~Dietz, M.~Multrus, V.~Eksler\emph{,~et~al.}, ``Overview of the evs codec architecture,'' in \emph{2015 IEEE International Conference on Acoustics, Speech and Signal Processing (ICASSP)}.\hskip 1em plus 0.5em minus 0.4em\relax IEEE, 2015, pp. 5698--5702.

\bibitem{valin2016high}
J.-M. Valin, G.~Maxwell, T.~B. Terriberry\emph{,~et~al.}, ``High-quality, low-delay music coding in the opus codec,'' \emph{arXiv preprint arXiv:1602.04845}, 2016.

\bibitem{klejsa2019high}
J.~Klejsa, P.~Hedelin, C.~Zhou\emph{,~et~al.}, ``High-quality speech coding with sample rnn,'' in \emph{ICASSP 2019-2019 IEEE International Conference on Acoustics, Speech and Signal Processing (ICASSP)}.\hskip 1em plus 0.5em minus 0.4em\relax IEEE, 2019, pp. 7155--7159.

\bibitem{kumar2024high}
R.~Kumar, P.~Seetharaman, A.~Luebs\emph{,~et~al.}, ``High-fidelity audio compression with improved rvqgan,'' \emph{Advances in Neural Information Processing Systems}, vol.~36, 2024.

\bibitem{yang2024generative}
H.~Yang, I.~Jang, and M.~Kim, ``Generative de-quantization for neural speech codec via latent diffusion,'' in \emph{ICASSP 2024-2024 IEEE International Conference on Acoustics, Speech and Signal Processing (ICASSP)}.\hskip 1em plus 0.5em minus 0.4em\relax IEEE, 2024, pp. 1251--1255.

\bibitem{ju2024naturalspeech}
Z.~Ju, Y.~Wang, K.~Shen\emph{,~et~al.}, ``Naturalspeech 3: Zero-shot speech synthesis with factorized codec and diffusion models,'' \emph{arXiv preprint arXiv:2403.03100}, 2024.

\bibitem{san2024discrete}
R.~San~Roman, Y.~Adi, A.~Deleforge\emph{,~et~al.}, ``From discrete tokens to high-fidelity audio using multi-band diffusion,'' \emph{Advances in Neural Information Processing Systems}, vol.~36, 2024.

\bibitem{ho2020denoising}
J.~Ho, A.~Jain, and P.~Abbeel, ``Denoising diffusion probabilistic models,'' \emph{Advances in neural information processing systems}, vol.~33, pp. 6840--6851, 2020.

\bibitem{rombach2022high}
R.~Rombach, A.~Blattmann, D.~Lorenz\emph{,~et~al.}, ``High-resolution image synthesis with latent diffusion models,'' in \emph{Proceedings of the IEEE/CVF conference on computer vision and pattern recognition}, 2022, pp. 10\,684--10\,695.

\bibitem{ren2023fewer}
Y.~Ren, T.~Wang, J.~Yi\emph{,~et~al.}, ``Fewer-token neural speech codec with time-invariant codes,'' \emph{arXiv preprint arXiv:2310.00014}, 2023.

\bibitem{li2024singlecodec}
\BIBentryALTinterwordspacing
H.~Li, L.~Xue, H.~Guo\emph{,~et~al.}, ``Single-codec: Single-codebook speech codec towards high-performance speech generation,'' 2024. [Online]. Available: \url{https://arxiv.org/abs/2406.07422}
\BIBentrySTDinterwordspacing

\bibitem{lee2023bigvgan}
\BIBentryALTinterwordspacing
S.~gil Lee, W.~Ping, B.~Ginsburg\emph{,~et~al.}, ``Bigvgan: A universal neural vocoder with large-scale training,'' 2023. [Online]. Available: \url{https://arxiv.org/abs/2206.04658}
\BIBentrySTDinterwordspacing

\bibitem{wang2004image}
Z.~Wang, A.~C. Bovik, H.~R. Sheikh\emph{,~et~al.}, ``Image quality assessment: from error visibility to structural similarity,'' \emph{IEEE transactions on image processing}, vol.~13, no.~4, pp. 600--612, 2004.

\bibitem{wu2023audiodec}
Y.-C. Wu, I.~D. Gebru, D.~Markovi{\'c}\emph{,~et~al.}, ``Audiodec: An open-source streaming high-fidelity neural audio codec,'' in \emph{ICASSP 2023-2023 IEEE International Conference on Acoustics, Speech and Signal Processing (ICASSP)}.\hskip 1em plus 0.5em minus 0.4em\relax IEEE, 2023, pp. 1--5.

\bibitem{ai2024apcodec}
Y.~Ai, X.-H. Jiang, Y.-X. Lu\emph{,~et~al.}, ``Apcodec: A neural audio codec with parallel amplitude and phase spectrum encoding and decoding,'' \emph{arXiv preprint arXiv:2402.10533}, 2024.

\bibitem{rix2001perceptual}
A.~W. Rix, J.~G. Beerends, M.~P. Hollier\emph{,~et~al.}, ``Perceptual evaluation of speech quality (pesq)-a new method for speech quality assessment of telephone networks and codecs,'' in \emph{2001 IEEE international conference on acoustics, speech, and signal processing. Proceedings (Cat. No. 01CH37221)}, vol.~2.\hskip 1em plus 0.5em minus 0.4em\relax IEEE, 2001, pp. 749--752.

\bibitem{taal2010short}
C.~H. Taal, R.~C. Hendriks, R.~Heusdens\emph{,~et~al.}, ``A short-time objective intelligibility measure for time-frequency weighted noisy speech,'' in \emph{2010 IEEE international conference on acoustics, speech and signal processing}.\hskip 1em plus 0.5em minus 0.4em\relax IEEE, 2010, pp. 4214--4217.

\bibitem{chinen2020visqol}
M.~Chinen, F.~S. Lim, J.~Skoglund\emph{,~et~al.}, ``Visqol v3: An open source production ready objective speech and audio metric,'' in \emph{2020 twelfth international conference on quality of multimedia experience (QoMEX)}.\hskip 1em plus 0.5em minus 0.4em\relax IEEE, 2020, pp. 1--6.

\bibitem{saeki2022utmos}
T.~Saeki, D.~Xin, W.~Nakata\emph{,~et~al.}, ``Utmos: Utokyo-sarulab system for voicemos challenge 2022,'' \emph{arXiv preprint arXiv:2204.02152}, 2022.

\bibitem{van2017neural}
A.~Van Den~Oord, O.~Vinyals\emph{,~et~al.}, ``Neural discrete representation learning,'' \emph{Advances in neural information processing systems}, vol.~30, 2017.

\bibitem{van2016wavenet}
A.~Van Den~Oord, S.~Dieleman, H.~Zen\emph{,~et~al.}, ``Wavenet: A generative model for raw audio,'' \emph{arXiv preprint arXiv:1609.03499}, vol.~12, 2016.

\bibitem{jiang2023disentangled}
X.~Jiang, X.~Peng, Y.~Zhang\emph{,~et~al.}, ``Disentangled feature learning for real-time neural speech coding,'' in \emph{ICASSP 2023-2023 IEEE International Conference on Acoustics, Speech and Signal Processing (ICASSP)}.\hskip 1em plus 0.5em minus 0.4em\relax IEEE, 2023, pp. 1--5.

\bibitem{chen2021again}
Y.-H. Chen, D.-Y. Wu, T.-H. Wu\emph{,~et~al.}, ``Again-vc: A one-shot voice conversion using activation guidance and adaptive instance normalization,'' in \emph{ICASSP 2021-2021 IEEE International Conference on Acoustics, Speech and Signal Processing (ICASSP)}.\hskip 1em plus 0.5em minus 0.4em\relax IEEE, 2021, pp. 5954--5958.

\bibitem{9747763}
C.~Ho~Chan, K.~Qian, Y.~Zhang\emph{,~et~al.}, ``Speechsplit2.0: Unsupervised speech disentanglement for voice conversion without tuning autoencoder bottlenecks,'' in \emph{ICASSP 2022 - 2022 IEEE International Conference on Acoustics, Speech and Signal Processing (ICASSP)}, 2022, pp. 6332--6336.

\bibitem{9746973}
V.~A. Trinh and S.~Braun, ``Unsupervised speech enhancement with speech recognition embedding and disentanglement losses,'' in \emph{ICASSP 2022 - 2022 IEEE International Conference on Acoustics, Speech and Signal Processing (ICASSP)}, 2022, pp. 391--395.

\bibitem{pan2023gemo}
Y.~Pan, Y.~Hu, Y.~Yang\emph{,~et~al.}, ``Gemo-clap: Gender-attribute-enhanced contrastive language-audio pretraining for speech emotion recognition,'' \emph{arXiv preprint arXiv:2306.07848}, 2023.

\bibitem{yao2023promptvc}
J.~Yao, Y.~Yang, Y.~Lei\emph{,~et~al.}, ``Promptvc: Flexible stylistic voice conversion in latent space driven by natural language prompts,'' \emph{arXiv preprint arXiv:2309.09262}, 2023.

\bibitem{pan2024gmp}
Y.~Pan, Y.~Yang, H.~Lu\emph{,~et~al.}, ``Gmp-atl: Gender-augmented multi-scale pseudo-label enhanced adaptive transfer learning for speech emotion recognition via hubert,'' \emph{arXiv preprint arXiv:2405.02151}, 2024.

\bibitem{wang2021vqmivc}
D.~Wang, L.~Deng, Y.~T. Yeung\emph{,~et~al.}, ``Vqmivc: Vector quantization and mutual information-based unsupervised speech representation disentanglement for one-shot voice conversion,'' \emph{arXiv preprint arXiv:2106.10132}, 2021.

\bibitem{pan2023msac}
Y.~Pan, ``Msac: Multiple speech attribute control method for speech emotion recognition,'' \emph{arXiv preprint arXiv:2308.04025}, 2023.

\bibitem{wang2023cam}
H.~Wang, S.~Zheng, Y.~Chen\emph{,~et~al.}, ``Cam++: A fast and efficient network for speaker verification using context-aware masking,'' \emph{arXiv preprint arXiv:2303.00332}, 2023.

\bibitem{pan2024ctefm}
Y.~Pan, Y.~Yang, J.~Yao\emph{,~et~al.}, ``Ctefm-vc: Zero-shot voice conversion based on content-aware timbre ensemble modeling and flow matching,'' \emph{arXiv preprint arXiv:2411.02026}, 2024.

\bibitem{vaswani2017attention}
A.~Vaswani, N.~Shazeer, N.~Parmar\emph{,~et~al.}, ``Attention is all you need,'' \emph{Advances in neural information processing systems}, vol.~30, 2017.

\bibitem{yang2023hybridformer}
Y.~Yang, Y.~Pan, J.~Yin\emph{,~et~al.}, ``Hybridformer: Improving squeezeformer with hybrid attention and nsr mechanism,'' in \emph{ICASSP 2023-2023 IEEE International Conference on Acoustics, Speech and Signal Processing (ICASSP)}.\hskip 1em plus 0.5em minus 0.4em\relax IEEE, 2023, pp. 1--5.

\bibitem{kong2020hifi}
J.~Kong, J.~Kim, and J.~Bae, ``Hifi-gan: Generative adversarial networks for efficient and high fidelity speech synthesis,'' \emph{Advances in neural information processing systems}, vol.~33, pp. 17\,022--17\,033, 2020.

\bibitem{zen2019libritts}
H.~Zen, V.~Dang, R.~Clark\emph{,~et~al.}, ``Libritts: A corpus derived from librispeech for text-to-speech,'' \emph{arXiv preprint arXiv:1904.02882}, 2019.

\bibitem{kingma2014adam}
D.~P. Kingma and J.~Ba, ``Adam: A method for stochastic optimization,'' \emph{arXiv preprint arXiv:1412.6980}, 2014.

\end{thebibliography}

\newpage

\end{document}